\documentclass[twocolumn,aps,prb,ltxgrid]{revtex4}
\usepackage{epsfig}
\usepackage{graphicx}
\usepackage{amsbsy,amssymb,amsmath,bm,ulem}

\usepackage[usenames]{color}

\begin{document}

\title{Flux Penetration in Superconducting Strip with Edge-Indentation}
\author{J. I. Vestg\aa rden, D. V. Shantsev, Y. M. Galperin and T. H. Johansen}
\affiliation{Department of Physics and Center for Advanced
Materials and Nanotechnology,    University of Oslo, P. O. Box
1048 Blindern, 0316 Oslo, Norway}

\begin{abstract}

~\vspace{3mm}

The flux penetration near a semicircular indentation 
at the edge of a thin superconducting strip placed in
a transverse magnetic field is investigated.
The flux front distortion due to the indentation
is calculated numerically by solving 
the Maxwell equations with a highly nonlinear $E(j)$ law.
We find that the excess penetration, $\Delta$, can be significantly
($\sim$~50~\%) larger than the indentation radius $r_0$,
in contrast to a bulk supercondutor in the critical state
where $\Delta=r_0$. 
It is also shown that  the flux creep tends to smoothen the flux front, i.e. reduce $\Delta$.
The results are in very good agreement with 
magneto-optical studies of flux penetration into an YBa$_2$Cu$_3$O$_x$ film
having an edge defect.

\end{abstract}

\maketitle

\section{Introduction}

Magnetic field penetrates type-II superconductors as a set of quantized
flux lines -- vortices. Macroscopically, the vortex matter can be
considered as a ``flux liquid''. An important feature of this matter
is pinning of vortices leading to
zero electrical resistance at zero temperature. The pinning results in 
a non-uniform distribution of
magnetic flux forming a
\textit{critical state}.  The critical state determines the
macroscopic properties,  e.g. the
maximum current density and magnetic susceptibility,
that are important for applications. According to the
critical state model,\cite{bean64} at any point of the sample the local value
of the electrical current density is equal to its critical value, 
$j_c$, for a given magnetic field and temperature.

An interesting property of the critical state is that 
{\em local} material defects
affect the field and current distributions on a {\em global}
scale. For example, 
even a small non-superconducting cavity or 
an edge indentation create sample-spanning discontinuity lines
where the current flow direction changes abruptly.\cite{CCS} 
At a non-zero temperature, the critical state is relaxed due to 
flux creep that is conventionally described by a highly
nonlinear current-voltage curve, $E \propto j^n$ 
where $n\gg 1$ and $E$ is electric field.
Nevertheless, the same tendency persists: a small cavity 
of size $\ell$ in a bulk superconductor
perturbs the field distributions
on a much larger scale of $\sim n\ell$.\cite{gurevich00}
Many applications of superconductors are based on thin films where   
this tendency must be even stronger since 
the relation between the magnetic field and current is 
{\em nonlocal}.\cite{brandt93}
Usually this leads to poorer performance of superconducting devices
whose global properties are deteriorated by numerous natural defects
blocking the current flow.
However, the same tendency can help control the flux motion on a global
scale by patterning the superconductor with arrays of small holes designed, e.g.,
to guide the flux in a particular direction.\cite{wordenweber04,yurchenko06}
  
Surprisingly, a quantitative understanding 
on how a single local defect affects the flux penetration
into a superconducting film is still rather poor.
Even the simple case of an infinitely long thin strip 
with a semicircular edge-indentation is not solved.     
For a bulk superconductor in the critical state such an indentation creates 
an excess flux penetration exactly equal to the indentation radius.\cite{CCS}
However the nonlocal electrodynamics in thin films and hence
the presence of Meissner currents in the flux-free regions
make the picture much more complicated.\cite{eisenmenger01} 
It has been observed using magneto-optic imaging that the excess flux penetration 
in films can significantly exceed the size of the 
indentation it originates from.\cite{schuster96}
The physical mechanism behind this enhancement is however not yet understood.
It could be related to the effect of thin-film geometry, to the flux creep or
to thermal instabilities nucleated at the indentation. 

This work aims to clarify this question by presenting
a detailed study of flux penetration into a strip with a semicircular 
edge-indentation in the flux creep regime.
We determine how the excess penetration 
depends on the size of the indentation, 
the applied magnetic field and the creep exponent $n$.

\section{Model}

\begin{figure}[t]
  \centering
  \epsfig{file=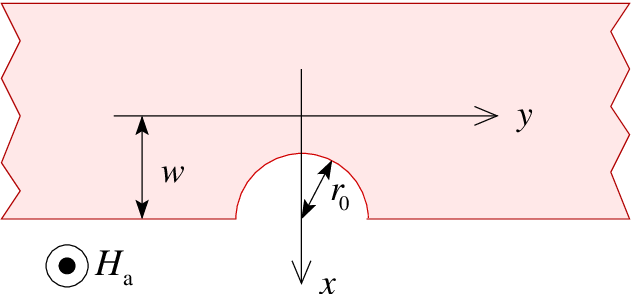,width=8.0cm} 
  \caption{
    A superconducting strip of width $2w$ with
    a semicircular indentation of radius $r_0$ in
    transverse field $H_a$.}
  \label{fig:sample}
\end{figure}

Consider a thin superconducting strip
of thickness $d$ 
placed in a transverse magnetic field. 
The strip is infinite in the $y$-direction, has the width $2w\gg d$ in the $x$-direction
and a semicircular indentation with radius $r_0$ at the edge, see Fig.~\ref{fig:sample}.
The flux dynamics in the creep regime is
conventionally described using  
a local relation between electric field $\mathbf E$ and 
current density $\mathbf j$,\cite{brandt95-prl,zeldov90,brandt95}
\begin{equation}
  \mathbf E = \rho \mathbf j\, ,
\end{equation}
with a highly nonlinear resistivity 
\begin{equation}
  \rho=\rho_0\left(j/j_c\right)^{n-1}\,  ,
  \label{material-law}
\end{equation}
which does not explicitly depend on the magnetic induction $B$. Here 
$\rho_0$ is a constant, $j_c$ is the critical current density,
while $n$ is the creep exponent, $n\gg 1$. This exponent can be related to the activation
energy $U$ for thermal depinning as $n\sim U/kT$.
Hence, large $n$ means small creep, and the Bean critical state model\cite{bean64} 
is regained in the limit $n\to\infty$.

For numerical simulations 
of flux penetration into the strip 
we use the formalism developed by
Brandt\cite{brandt92,brandt93,schuster95,brandt95-prl,brandt95,mints96,brandt97,brandt01,brandt05}
that can be applied to thin type-II
superconductors of various shapes. 
For a thin superconductor, it is appropriate to look 
at length scales larger than the thickness $d$, and introduce a
sheet current 
$$\mathbf{J}(\mathbf r)=\int_{-d/2}^{d/2} dz~\mathbf{j}(\mathbf r,z)\, ,
$$
where  $\mathbf r=(x,y)$ are in-plane coordinates.
Due to the current conservation, $\nabla\cdot\mathbf J=0$, the sheet current
can be expressed through a scalar function $g(\mathbf r)$ as
\begin{equation}
  \label{eq:cd001}
  \mathbf J = \nabla \times \hat z g
\end{equation}
where $g$ has the interpretation of 
the local magnetization.\cite{brandt95-prl} 
Substituting the current from Eq.~(\ref{eq:cd001})
into the Biot-Savart law one arrives at a non-local relation 
between $B_z$ and $g$,
\begin{equation}
  B_z(\mathbf r,z)
  = \mu_0H_a+
  \int_{A} d^2 r'\, Q(\mathbf r, \mathbf r',z)\, g(\mathbf r') \,  .
  \label{hfromg}
\end{equation}
Here $H_a$ is the applied field and $A$ is the sample area.
The integral kernel 
is equal to the field of a dipole of unit strength,
\begin{equation}
 Q(\mathbf r,\mathbf r', z) = \frac{\mu_0}{4\pi}
  \frac{2z^2-(\mathbf r - \mathbf r')^2}
       {\left[z^2+(\mathbf r-\mathbf r')^2\right]^{5/2}}\,  .
       \label{kernel1}
\end{equation}
The integral Eq.~\eqref{hfromg} with kernel Eq.~\eqref{kernel1} is
divergent at $\mathbf r \to \mathbf r'$ and  $z\to 0$. In a numerical
procedure, the divergence can be handled in three ways: (i) by keeping
a finite $z$ during the calculation;\cite{loerincz04} 
(ii) by working in the Fourier space;\cite{brandt95} (iii)
by converting the integral to a matrix form and 
using the flux conservation to determine the diagonal
elements.\cite{brandt05,brandt92,brandt01}  
Here we use the third method. 
Since, for $H_a=0$, the total flux through the  $z = 0$ plane is zero, the kernel 
should have the property $\int d^2r\, Q(\mathbf r,\mathbf r',0)=0$. 
This yields
\begin{equation}
  \frac{1}{\mu_0}B_z(\mathbf r)
  =  H_a+
  g(\mathbf r) C(\mathbf r)
  -
  \int_{A} \frac{d^2r'}{4\pi}
  \frac{g(\mathbf r')- g(\mathbf r)}
       {|\mathbf r- \mathbf r'|^3}
       ,
       \label{biot-savart-2}
\end{equation}
where the scalar function $C$ is an integral over the area \textit{outside} the
superconductor 
\begin{equation} 
  \label{C}
  C(\mathbf r) 
  =  
  \int_{\text{outside}} 
  \frac{dr'^2}
       {4\pi |\mathbf r - \mathbf r'|^3}\,      .
\end{equation}
For a uniform strip of width $2w$ it yields 
\begin{equation}
  C_\text{strip}(x)=\frac{1}{\pi}\frac{w}{w^2-x^2}.
\end{equation}
In addition, the indentation gives a contribution from the semicircle, which is
calculated numerically from~Eq.~\eqref{C}. 
\begin{figure}[t]
  \centering
  \epsfig{file=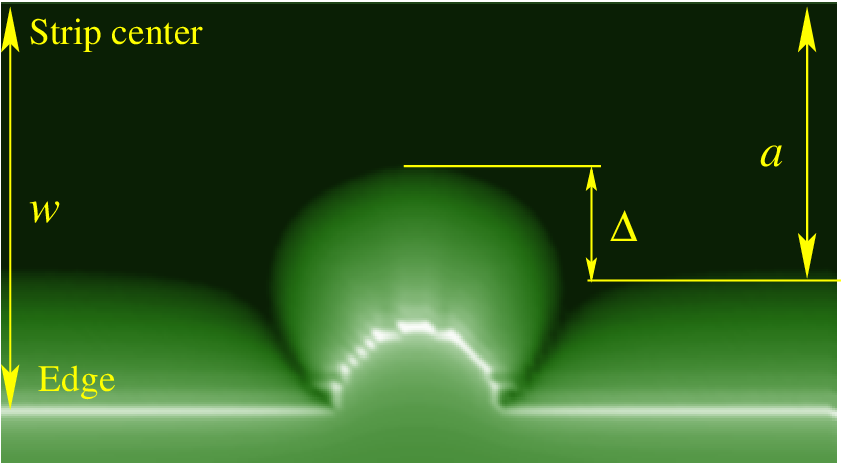,width=8.0cm} 
  \caption{The simulated flux density map in a strip of width $2w$,
    with a semicircular indentation of radius $r_0=0.2w$, in applied field $H_a=0.3 J_c$, $n=19$, and
    ramped with a rate $\mu_0\dot H_a=\rho_0J_c/wd$.
    Note that $\Delta$ is not equal to $r_0$.}
  \label{fig:mo-sample}
  \label{fig:H}
\end{figure}

In the following we use an equidistant square grid and ascribe the same area $s$
to each grid point. 
The discrete version of the kernel then acquires the
form\cite{brandt05}  
\begin{equation}
  \frac{Q_{ij}}{\mu_0}
  =
    \delta_{ij}\left(\frac{C_i}{s}+\sum_l  q_{il}\right)
    - q_{ij}    \, ,
  \label{kernel-discrete}
\end{equation}
where  
$q_{ij}= 1/4\pi|\mathbf r_i - \mathbf r_j|^3$ for $i\neq j$ and
$q_{ii}=0$.  
All elements of the discrete kernel Eq.~\eqref{kernel-discrete} are 
nondivergent and the flux conservation, $\int d^2r \, B(\mathbf r)=0$, is guaranteed.
Relating the magnetic field and the current by the Faraday's  
law, and using the inverted Biot-Savart law one obtains the dynamic
equation for the local magnetization:
\begin{equation}
  \dot g(\mathbf r,t) 
  =
  \int_A d^2 r'\, Q^{-1}(\mathbf r,\mathbf r')\, \left[\hat{f}
    g(\mathbf{r}',t)
-\dot H_a(t)\right] 
  ,
  \label{dynamics}
\end{equation}
where 
$$\hat{f} g \equiv \nabla\cdot(\rho\nabla g) / d\mu_0\, .$$ 
For discrete formulation of the problem the inverse kernel 
$Q^{-1}$ 
is just the inverse of the matrix Eq.~\eqref{kernel-discrete}, hence 
the matrix must be calculated and inverted only once.

\begin{figure}[t]
  $\begin{array}{c}
  \epsfig{file=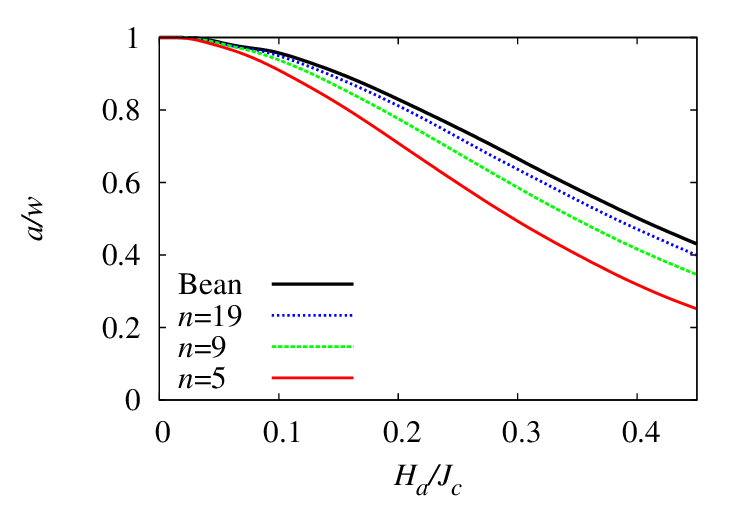,width=9cm} \\
  \end{array}$
  \caption{Evolution of flux penetration depth $a$ 
    as the applied field is ramped with a constant rate
    $\mu_0\dot H_a=\rho_0J_c/wd$. Stronger flux creep, i.e. smaller $n$, 
    leads to deeper penetration. The Bean limit is $n=101$.
  }
  \label{fig:fluxfront}
\end{figure}

\begin{figure}[t]
  $\begin{array}{c}
    \epsfig{file=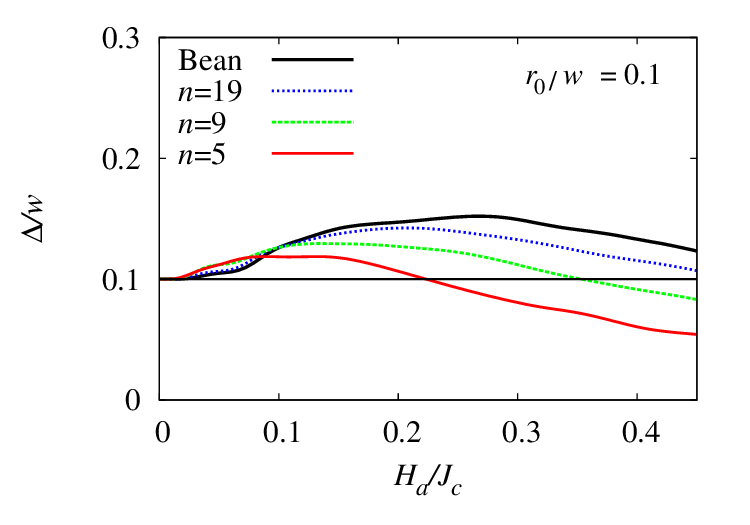,width=8cm} \\
    \epsfig{file=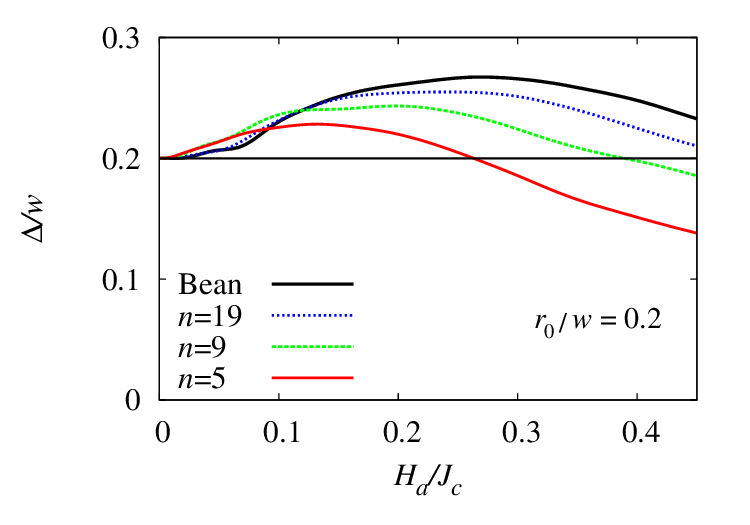,width=8cm} \\
  \end{array}$
  \caption{Evolution of the indentation-induced excess penetration, $\Delta$,
    as a function of applied field. 
    The two panels correspond to
    different indentation radii, $r_0/w=$0.1 and~0.2, respectively; $\mu_0\dot H_a=\rho_0J_c/wd$. 
	}
  \label{fig:distortion}
\end{figure}
%

\section{Results and Discussion}

\paragraph{Magnetic field and current}

The simulations were performed by ramping the applied field
at a constant rate~$\mu_0\dot H_a=\rho_0J_c/wd$, starting at zero field
and a flux-free strip. The flux penetrates
from the edges forming well-defined flux fronts that move
towards the strip center as the applied field increases.
Shown in Fig.~\ref{fig:H} is a typical result of the flux 
density distribution presented as seen in a 
magneto-optical image, i.e., the image brightness 
represents the magnitude of the perpendicular magnetic field.
The sample edge is seen as a bright line,
i.e., the flux density is highest at the edge.

Far from the indentation the flux penetration front is straight,
and leaves a fraction $a/w$ of the strip in the flux-free Meissner state,
seen here as a black region. The penetration of this straight front versus applied field is 
shown in Fig.~\ref{fig:fluxfront} for different values of the creep exponent.
For large $n$ the simulations approach the Bean-model result,\cite{brandt93}
$a_\text{Bean}=w/\cosh(\pi H_a/J_c)$,
while for smaller $n$, i.e., stronger flux creep, the penetration is deeper,
all as expected for a strip with straight edges.

\begin{figure}[t]
  $
  \begin{array}{l}
  \begin{array}{cc}
  \epsfig{file=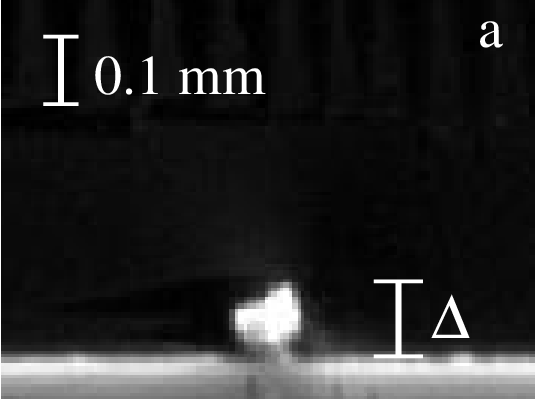,height=2.4cm} &
  \epsfig{file=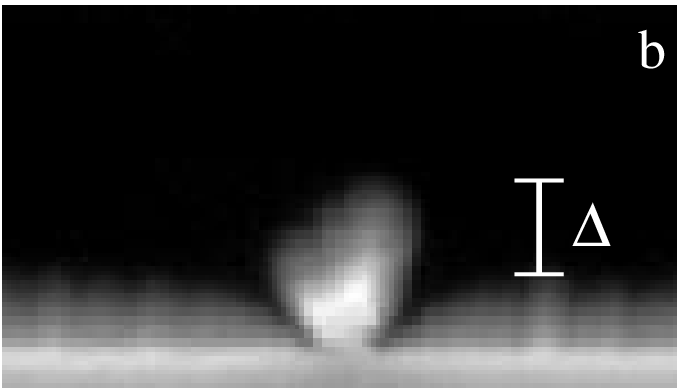,height=2.4cm} 
  \end{array} \\
  \epsfig{file=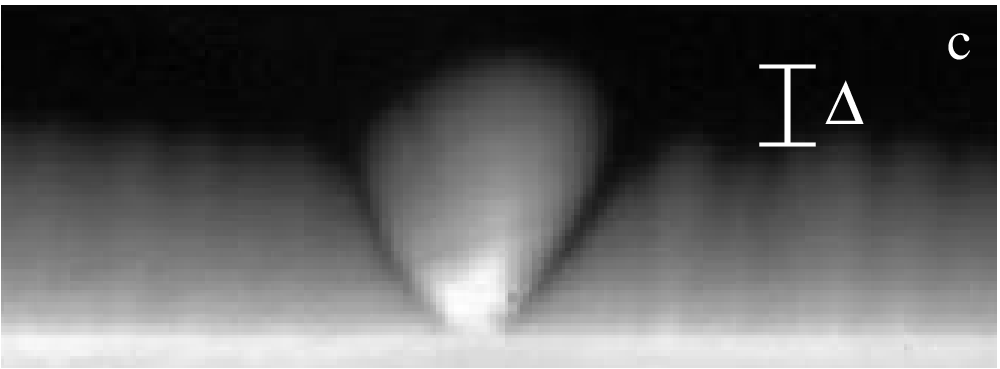,height=2.4cm}
  \end{array}
  $
  
  \caption{Magneto-optical images of 
    flux penetration into an 
    YBa$_2$Cu$_3$O$_x$ strip with 
    a defect at the
    edge. Only the lower half of the strip is shown.
    In (a), (b) and (c) the applied fields was 3, 17 and 36~mT,
    respectively.
    The excess flux penetration, $\Delta$, is maximal at 
    the intermediate field,
    in agreement with simulations.
} 
  \label{fig:ybco}
\end{figure}

Near the indentation the flux penetration largely follows the circular shape.
At both sides of the indentation there are dark regions of reduced flux density. 
As penetration gets deeper these will become narrow $d$-lines,
where the current stream lines make sharp turns.\cite{CCS}
In the Bean limit
$n\to\infty$, the $d$-lines of semicircular indentations have parabolic shape.
With finite $n$ the parabolic shape is only approximated.
However, the main effect of the indentation is that it pushes magnetic field 
deeper into the sample. In order to quantify this we
define the excess penetration $\Delta$ as the difference between the deepest 
penetration and the penetration far away from the indentation.
Fig.~\ref{fig:distortion} shows how $\Delta$ evolves with increasing $H_a$. 
Evidently, the excess penetration is not equal to 
the indentation radius, $r_0$, as in the case of the bulk Bean model.\cite{CCS,schuster94,gurevich00}
Moreover, $\Delta$ turns out to be field-dependent. Initially, $\Delta$
increases, then reaches a maximum followed by a decrease at larger $H_a$. 
This surprising non-monotonous behaviour is supported by magneto-optical 
measurements of the flux penetration in a uniform YBa$_2$Cu$_3$O$_x$ film
containing an edge defect, see Fig.~\ref{fig:ybco}.  
The film was shaped as a strip of half-width $w=0.4$~mm,
and the figure shows the flux distribution at 25 K for 3 different
applied fields. In (a) the field was very small, $\mu_0 H_a =  3 $~mT,
creating negligible penetration so that the actual shape of the defect 
appears in the image as the bright "bay area" inside the strip. 
In this state the excess penetration is equal to the depth of the defect,
 and measures  $\Delta = 80~\mu$m.  In (b) and (c) the applied field 
is 17~mT and 36~mT, respectively, and the corresponding excess 
penetration is  $\Delta = 115~\mu$m and 100 $\mu$m. This gives 
for $\Delta/w = 0.20, 0.29$ and 0.25, demonstrating an excess 
penetration that exceeds the depth of the indentation by 
nearly 40 \%, in very good agreement with the Bean model 
results plotted in Fig.~\ref{fig:distortion}. 

The Fig.~\ref{fig:distortion} includes the behaviour of $\Delta/w$ for two different $r_0/w$.
Comparing the two panels we see that larger indentations produce a larger $\Delta$.
However, the relative excess penetration, $\Delta/r_0$ is 
larger for the {\em smaller} indentation. The excess penetration can exceed 
the indentation depth by almost 50 \%
for $r_0=0.1w$ and large values of $n$.
For smaller values of the creep exponent
one always finds smaller $\Delta$, implying that
creep tends to smoothen perturbations in the flux front.\cite{gurevich00}

\begin{figure}[t]
  $\begin{array}{ccc}
  \epsfig{file=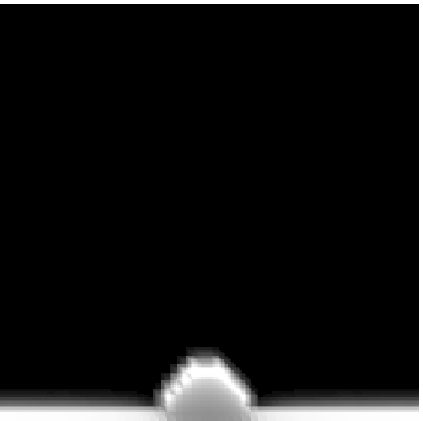,width=2.4cm} &
  \epsfig{file=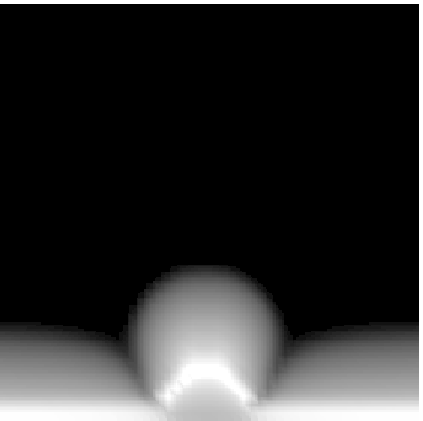,width=2.4cm} &
  \epsfig{file=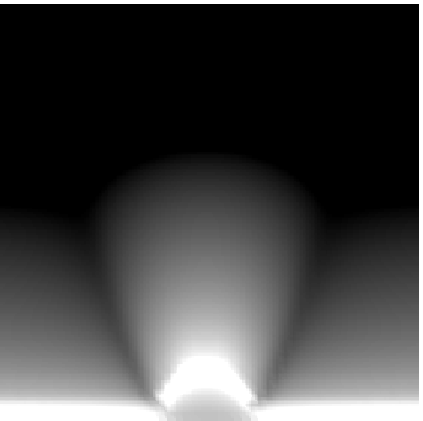,width=2.4cm} \\
  \epsfig{file=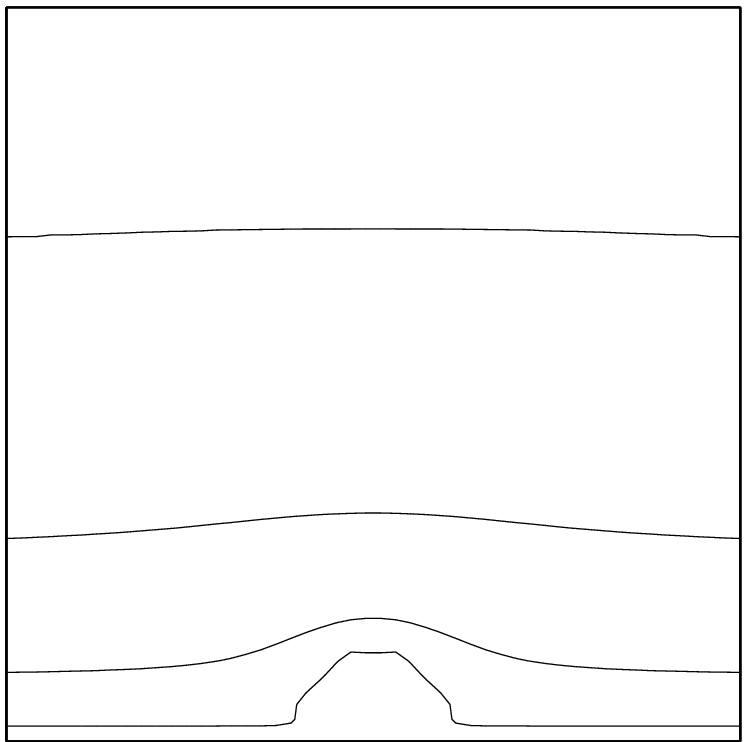,width=2.4cm} &
  \epsfig{file=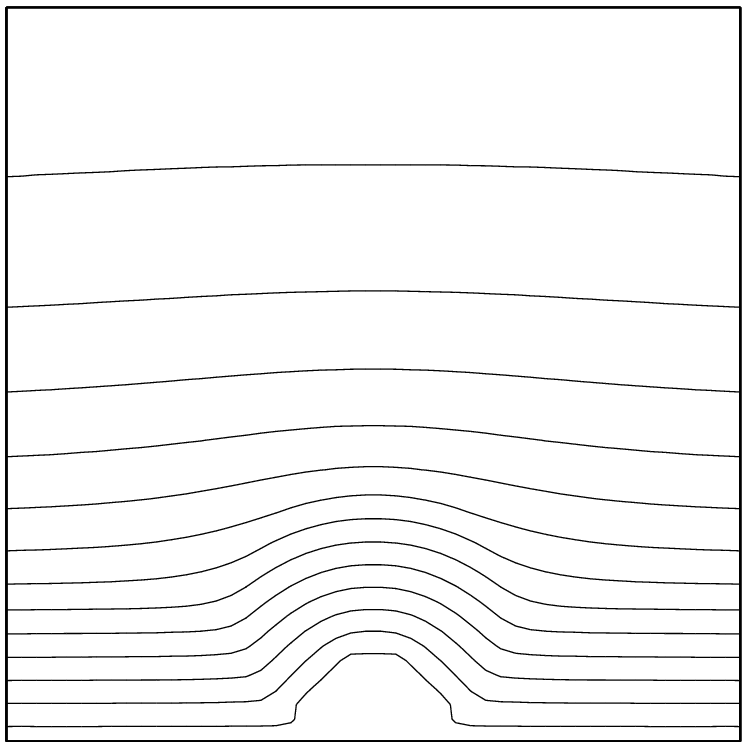,width=2.4cm} &
  \epsfig{file=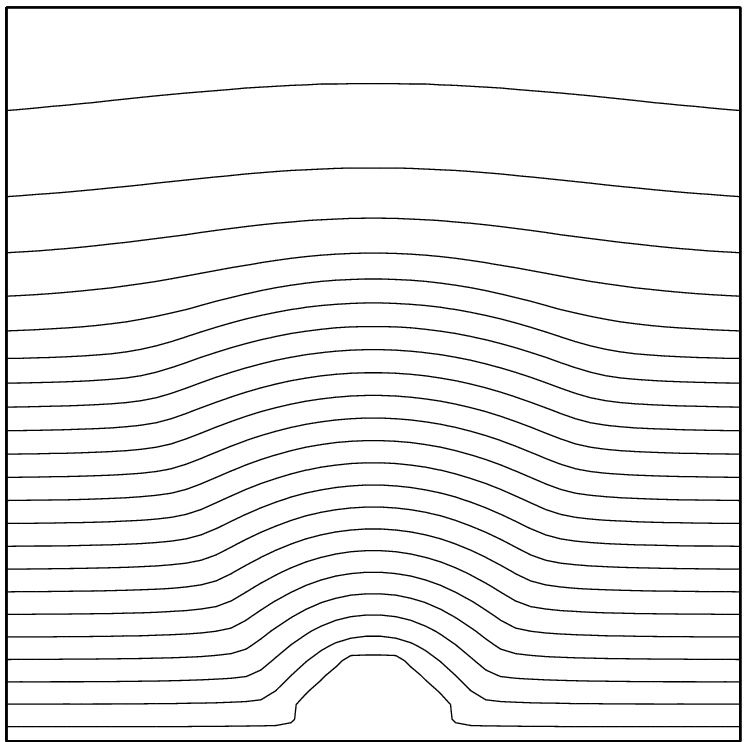,width=2.4cm}
  \end{array}$
  \caption{
    Simulated flux distributions (top) and current stream lines
    (bottom) in an increasing applied field, where
    $r_0=0.1w$ and the other 
    parameters the same as in Fig.~\ref{fig:H}.
    From left, the values of $H_a$
    are $0.05J_c$, $0.2J_c$, and $0.4J_c$ 
    with corresponding values of $a$ 
    $0.98w$, $0.8w$, and $0.5w$.
} 
  \label{fig:H+J}
\end{figure}

Our results demonstrate that an indentation in a thin film 
affects the flux distribution in a stronger and more complex way
than it does in bulk superconductors. 
This must be due to the non-local electrodynamics of thin films, and
in particular due to the presence of Meissner currents in the
flux free regions. These Meissner currents 
do not make the same sharp turns as the critical currents
in the flux penetrated region, see Fig.~\ref{fig:H+J} and also Refs.~\onlinecite{brandt95-prl,brandt95}. 
As a result, the Meissner currents concentrate in front of the indentation
where their density reaches $j_c$ and hence leads to even deeper flux penetration.
This is why the flux front near the indentation advances faster
than in the rest of the film.
This accelerated advancement eventually terminates 
when the penetration depth becomes comparable to the strip halfwidth.
The reason is simply that in the limit of full penetration 
all flux fronts reach the middle of the strip and hence $\Delta \to 0$.

\paragraph{Electric field}

The Lorentz force pushing magnetic flux is directed
perpendicular to the local current density.
Even a small indentation distorts the current stream lines  
over a large area, and hence significantly modifies the trajectories
of flux motion. 
In particular, all the flux arriving to the
fan-shaped region rooted at the indentation 
must have entered the sample through this indentation, see Fig.~\ref{fig:ybco}.
It creates a dramatic local enhancement of electric field 
since $E$ is a direct measure of the intensity of flux traffic.

Analytical solution for the electric field distribution 
around an indentation in thin films is not available. Therefore the
results obtained for the case of a slab are often utilized as
approximations also for films.\cite{mints96,gurevich00,gurevich01}
We will now analyze to what extent such estimates
are valid by comparing them with our simulation results for a strip.

In the fan-shaped region that originates from the semicircular indentation,
the electric field can be found by solving the Maxwell equation 
$\nabla\times \mathbf E=-\dot {\mathbf B}$
in cylindrical coordinates. 
Since the evolution of $B$-distribution is usually not
very far from the Bean model, one can assume 
$\dot B=\mu_0\dot H_a$, which leads to the solution\cite{mints96}
\begin{equation}
  E_1(x) = \frac{\mu_0\dot
    H_a}{2}\left[\frac{\left(w-a+r_0\right)^2}{w-|x|}-(w-|x|)\right] \label{eq:bulk001}
\end{equation}
for $|x| > a - r_0$ and zero for $|x| < a-r_0$. 
Far away from the indentation the solution of the same equation
in cartesian coordinates,
$\partial_x E=-\mu_0\dot H_a$, is 
\begin{equation}
  E_0(x) = \mu_0\dot H_a\left(|x|-a\right) \label{eq:bulk0012}
\end{equation}
for $|x|>a$ and zero for $|x|<a$. 
Note that the width $w$ enters Eq.~(\ref{eq:bulk001})  
only because of the  
specific choice of the $x$-coordinate, where the edge is located at $x=w$. 
Replacement $x \to x+w$ removes the $w$-dependence.

\begin{figure}[t]
  \epsfig{file=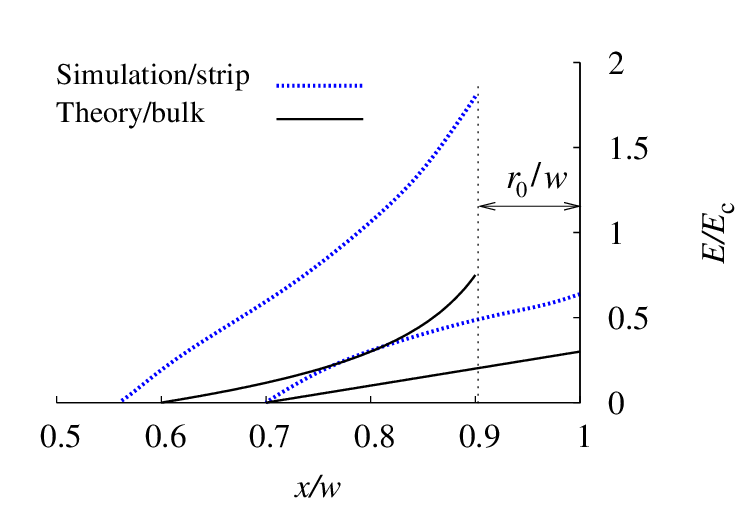,width=7cm} 
  \caption{Electric field profiles 
    across the indentation and far away from it. 
    Solid lines: Eqs.~\ref{eq:bulk001} and~\ref{eq:bulk0012} for a bulk
    superconductor. Dashed lines: simulations for a thin-film strip;
    the parameters are the same as for Fig.~\ref{fig:H} except 
    that $r_0=0.1w$ and $H_a/J_c=0.25$. $E_c=\rho_0J_c/d$.
    A strong field enhancement near the indentation is clearly seen. 
 }
  \label{fig:E-profile}
\end{figure}

Figure~\ref{fig:E-profile} compares $E_0(x)$ and $E_1(x)$ with
the simulated electric field profiles.
The quantitative agreement is poor, though
the shape of profiles (both across the
indentation and away from it) is fairly well reproduced, 
in agreement with Ref.~\onlinecite{schuster96}.
The expected enhancement of $E$ due to indentation is also obvious.
The formulas above predict 
the relative enhancement for the peak values
$E_1^{(\max)} / E_0^{(\max)} = (w-a)/2r_0+1$ for a bulk sample.
One can see from the plot that the effect of indentation 
is even stronger for thin films: 
the ratio $E_1^{(\max)} / E_0^{(\max)}$ is slightly higher and the
excess penetration is larger 
(the flux front here corresponds to the point where $E(x)=0$).

A locally enhanced electric field 
near edge indentations
and hence enhanced Joule heating
is predicted to 
facilitate nucleation of a thermal instability.\cite{mints96,gurevich01}
The instability in thin superconductors is usually observed in form of
macroscopic dendritic flux avalanches\cite{denisov05} or
macroscopic uniform flux jumps\cite{prozorov}.
However, a third scenario is also possible when 
a series of {\em microscopic} flux avalanches repeatedly take place
in the same region, each leading 
to a small advancement of the flux front.\cite{shantsev05}
It creates an additional front distortion since the avalanches
are expected to be larger and occur more
frequently at the indentation, where the local $E$
is maximal. 
Experimentally the individual avalanches can be very small,
and hence it is not easy to determine 
whether the thermal effects contribute to an observed front distortion.
To identify the penetration mechanism one can compare the observed
flux profiles with the simulations.
The maximal excess penetration due to non-thermal effects is found
to be 150~\% of the indentation radius 
for our parameters. Consequently, when the observed  
excess penetration is larger, the flux penetration probably occurs via
thermal micro-avalanches.

\section{Conclusions} 
We have numerically solved the Maxwell equations to describe
flux penetration into a thin superconducting strip with an edge-indentation and
analyzed the time evolution of flux front in an increasing applied field, $H_a$.
The excess penetration, $\Delta$, due to the indentation
is not equal to the indentation radius, $r_0$, in contrast to the well-known case of
a bulk superconductor in the Bean model. 
Three different mechanisms that influence the excess penetration were analyzed.
(i) The nonlocal electrodynamics in films leads to a characteristic $\Delta(H_a)$
dependence with a smooth peak. The ratio $\Delta/r_0$ at the peak
equals 1.5 when $r_0$ is 0.1 of the strip half-width and becomes
even larger for smaller $r_0$.
(ii) The flux creep always tends to smoothen the flux front and decrease 
the excess penetration.
(iii) Thermal flux avalanches are more likely to occur
at the indentation, which can increase the apparent front distortion.   
Our results can be very helpful in order to identify which of these three mechanisms
is the dominant one in a concrete experiment. \\

\acknowledgements
We thank C.~Romero and Ch.~Jooss for fruitful discussions. 
This work was supported financially by The Norwegian Research
Council, Grant No. 158518/431 (NANOMAT) and by FUNMAT@UIO. 

\appendix

\section{Numerical details}
The simulations are carried out on an equidistant square grid with $N\times N$
points, $x_m=w(2m+1)/N-w$ and $y_n=w(2n+1)/N-w$,
for $0\leq m,n<N$.
The system has two symmetries that must be incorporated in the kernel:
first, the periodic boundary, which means that we must add a 
mirror strips at $x<-w$ and $x>w$. Second, the symmetry around 
$x=0$. The latter means that we can work with half the kernel.\cite{brandt95} 
The simulations use a grid size of $N=100$, which means
that a $5000\times 5000$ matrix must be put in memory
and inverted. The memory consumption 
is the main limiting factor of the simulations.
The kernel is stable, so there is no
need for additional smoothening. 
For most exponents a pure power
law is used, but for the Bean limit, $n=101$, a cutoff on the
resistivity $\rho<\rho_{\text{max}}$ was necessary to ensure stability.
The flux front position was determined at every time step
and then smoothened as a function of time.
It allows the front position to be determined  
with an accuracy much better than the distance between two grid points.


\begin{thebibliography}{23}
\expandafter\ifx\csname natexlab\endcsname\relax\def\natexlab#1{#1}\fi
\expandafter\ifx\csname bibnamefont\endcsname\relax
  \def\bibnamefont#1{#1}\fi
\expandafter\ifx\csname bibfnamefont\endcsname\relax
  \def\bibfnamefont#1{#1}\fi
\expandafter\ifx\csname citenamefont\endcsname\relax
  \def\citenamefont#1{#1}\fi
\expandafter\ifx\csname url\endcsname\relax
  \def\url#1{\texttt{#1}}\fi
\expandafter\ifx\csname urlprefix\endcsname\relax\def\urlprefix{URL }\fi
\providecommand{\bibinfo}[2]{#2}
\providecommand{\eprint}[2][]{\url{#2}}

\bibitem[{\citenamefont{Bean}(1964)}]{bean64}
\bibinfo{author}{\bibfnamefont{C.~P.} \bibnamefont{Bean}},
  \bibinfo{journal}{Rev. Mod. Phys.} \textbf{\bibinfo{volume}{36}},
  \bibinfo{pages}{31} (\bibinfo{year}{1964}).

\bibitem[{\citenamefont{Campbell and Evetts}(1972)}]{CCS}
\bibinfo{author}{\bibfnamefont{A.~M.} \bibnamefont{Campbell}} \bibnamefont{and}
  \bibinfo{author}{\bibfnamefont{J.}~\bibnamefont{Evetts}},
  \emph{\bibinfo{title}{Critical Currents in Superconductors}}
  (\bibinfo{publisher}{Taylor and Francis LTD, London}, \bibinfo{year}{1972}).

\bibitem[{\citenamefont{Gurevich and Friesen}(2000)}]{gurevich00}
\bibinfo{author}{\bibfnamefont{A.}~\bibnamefont{Gurevich}} \bibnamefont{and}
  \bibinfo{author}{\bibfnamefont{M.}~\bibnamefont{Friesen}},
  \bibinfo{journal}{Phys. Rev. B} \textbf{\bibinfo{volume}{62}},
  \bibinfo{pages}{4004} (\bibinfo{year}{2000}).

\bibitem[{\citenamefont{Brandt and Indenbom}(1993)}]{brandt93}
\bibinfo{author}{\bibfnamefont{E.~H.} \bibnamefont{Brandt}} \bibnamefont{and}
  \bibinfo{author}{\bibfnamefont{M.}~\bibnamefont{Indenbom}},
  \bibinfo{journal}{Phys. Rev. B} \textbf{\bibinfo{volume}{48}},
  \bibinfo{pages}{12893} (\bibinfo{year}{1993}).

\bibitem[{\citenamefont{W{\" o}rdenweber et~al.}(2004)\citenamefont{W{\"
  o}rdenweber, Dymashevski, and Misko}}]{wordenweber04}
\bibinfo{author}{\bibfnamefont{R.}~\bibnamefont{W{\" o}rdenweber}},
  \bibinfo{author}{\bibfnamefont{P.}~\bibnamefont{Dymashevski}},
  \bibnamefont{and} \bibinfo{author}{\bibfnamefont{V.~R.} \bibnamefont{Misko}},
  \bibinfo{journal}{Phys. Rev. B} \textbf{\bibinfo{volume}{69}},
  \bibinfo{pages}{184504} (\bibinfo{year}{2004}).

\bibitem[{\citenamefont{Yurchenko et~al.}(2006)\citenamefont{Yurchenko, W{\"
  o}rdenweber, Galperin, Shantsev, Vestg{\aa}rden, and Johansen}}]{yurchenko06}
\bibinfo{author}{\bibfnamefont{V.~V.} \bibnamefont{Yurchenko}},
  \bibinfo{author}{\bibfnamefont{R.}~\bibnamefont{W{\" o}rdenweber}},
  \bibinfo{author}{\bibfnamefont{Y.~M.} \bibnamefont{Galperin}},
  \bibinfo{author}{\bibfnamefont{D.~V.} \bibnamefont{Shantsev}},
  \bibinfo{author}{\bibfnamefont{J.~I.} \bibnamefont{Vestg{\aa}rden}},
  \bibnamefont{and} \bibinfo{author}{\bibfnamefont{T.~H.}
  \bibnamefont{Johansen}}, \bibinfo{journal}{Physica C}
  \textbf{\bibinfo{volume}{437-438}}, \bibinfo{pages}{357}
  (\bibinfo{year}{2006}).

\bibitem[{\citenamefont{Eisenmenger et~al.}(2001)\citenamefont{Eisenmenger,
  Leiderer, Wallenhorst, and D{\" o}tsch}}]{eisenmenger01}
\bibinfo{author}{\bibfnamefont{J.}~\bibnamefont{Eisenmenger}},
  \bibinfo{author}{\bibfnamefont{P.}~\bibnamefont{Leiderer}},
  \bibinfo{author}{\bibfnamefont{M.}~\bibnamefont{Wallenhorst}},
  \bibnamefont{and} \bibinfo{author}{\bibfnamefont{H.}~\bibnamefont{D{\"
  o}tsch}}, \bibinfo{journal}{Phys. Rev. B} \textbf{\bibinfo{volume}{64}},
  \bibinfo{pages}{104503} (\bibinfo{year}{2001}).

\bibitem[{\citenamefont{Schuster et~al.}(1996)\citenamefont{Schuster, Kuhn, and
  Brandt}}]{schuster96}
\bibinfo{author}{\bibfnamefont{T.}~\bibnamefont{Schuster}},
  \bibinfo{author}{\bibfnamefont{H.}~\bibnamefont{Kuhn}}, \bibnamefont{and}
  \bibinfo{author}{\bibfnamefont{E.~H.} \bibnamefont{Brandt}},
  \bibinfo{journal}{Phys. Rev. B} \textbf{\bibinfo{volume}{54}},
  \bibinfo{pages}{3514} (\bibinfo{year}{1996}).

\bibitem[{\citenamefont{Brandt}(1995{\natexlab{a}})}]{brandt95-prl}
\bibinfo{author}{\bibfnamefont{E.~H.} \bibnamefont{Brandt}},
  \bibinfo{journal}{Phys. Rev. Lett.} \textbf{\bibinfo{volume}{74}},
  \bibinfo{pages}{3025} (\bibinfo{year}{1995}{\natexlab{a}}).

\bibitem[{\citenamefont{Zeldov et~al.}(1990)\citenamefont{Zeldov, Amer, Koren,
  Gupta, and Mc{E}lfresh}}]{zeldov90}
\bibinfo{author}{\bibfnamefont{E.}~\bibnamefont{Zeldov}},
  \bibinfo{author}{\bibfnamefont{N.~M.} \bibnamefont{Amer}},
  \bibinfo{author}{\bibfnamefont{G.}~\bibnamefont{Koren}},
  \bibinfo{author}{\bibfnamefont{A.}~\bibnamefont{Gupta}}, \bibnamefont{and}
  \bibinfo{author}{\bibfnamefont{M.~W.} \bibnamefont{Mc{E}lfresh}},
  \bibinfo{journal}{Appl. Phys. Lett.} \textbf{\bibinfo{volume}{56}},
  \bibinfo{pages}{680} (\bibinfo{year}{1990}).

\bibitem[{\citenamefont{Brandt}(1995{\natexlab{b}})}]{brandt95}
\bibinfo{author}{\bibfnamefont{E.~H.} \bibnamefont{Brandt}},
  \bibinfo{journal}{Phys. Rev. B} \textbf{\bibinfo{volume}{52}},
  \bibinfo{pages}{15442} (\bibinfo{year}{1995}{\natexlab{b}}).

\bibitem[{\citenamefont{Brandt}(1992)}]{brandt92}
\bibinfo{author}{\bibfnamefont{E.~H.} \bibnamefont{Brandt}},
  \bibinfo{journal}{Phys. Rev. B} \textbf{\bibinfo{volume}{46}},
  \bibinfo{pages}{8628} (\bibinfo{year}{1992}).

\bibitem[{\citenamefont{Schuster et~al.}(1995)\citenamefont{Schuster, Kuhn,
  Brandt, Indenbom, Kl{\" a}ser, M{\" u}ller{-}Vogt, Habermeier, Kronm{\"
  u}ller, and Forkl}}]{schuster95}
\bibinfo{author}{\bibfnamefont{T.}~\bibnamefont{Schuster}},
  \bibinfo{author}{\bibfnamefont{H.}~\bibnamefont{Kuhn}},
  \bibinfo{author}{\bibfnamefont{E.~H.} \bibnamefont{Brandt}},
  \bibinfo{author}{\bibfnamefont{M.~V.} \bibnamefont{Indenbom}},
  \bibinfo{author}{\bibfnamefont{M.}~\bibnamefont{Kl{\" a}ser}},
  \bibinfo{author}{\bibfnamefont{G.}~\bibnamefont{M{\" u}ller{-}Vogt}},
  \bibinfo{author}{\bibfnamefont{H.-U.} \bibnamefont{Habermeier}},
  \bibinfo{author}{\bibfnamefont{H.}~\bibnamefont{Kronm{\" u}ller}},
  \bibnamefont{and} \bibinfo{author}{\bibfnamefont{A.}~\bibnamefont{Forkl}},
  \bibinfo{journal}{Phys. Rev. B} \textbf{\bibinfo{volume}{52}},
  \bibinfo{pages}{10375} (\bibinfo{year}{1995}).

\bibitem[{\citenamefont{Mints and Brandt}(1996)}]{mints96}
\bibinfo{author}{\bibfnamefont{R.~G.} \bibnamefont{Mints}} \bibnamefont{and}
  \bibinfo{author}{\bibfnamefont{E.~H.} \bibnamefont{Brandt}},
  \bibinfo{journal}{Phys. Rev. B} \textbf{\bibinfo{volume}{54}},
  \bibinfo{pages}{12421} (\bibinfo{year}{1996}).

\bibitem[{\citenamefont{Brandt}(1997)}]{brandt97}
\bibinfo{author}{\bibfnamefont{E.~H.} \bibnamefont{Brandt}},
  \bibinfo{journal}{Phys. Rev. B} \textbf{\bibinfo{volume}{55}},
  \bibinfo{pages}{14513} (\bibinfo{year}{1997}).

\bibitem[{\citenamefont{Brandt}(2001)}]{brandt01}
\bibinfo{author}{\bibfnamefont{E.~H.} \bibnamefont{Brandt}},
  \bibinfo{journal}{Phys. Rev. B} \textbf{\bibinfo{volume}{64}},
  \bibinfo{pages}{024505} (\bibinfo{year}{2001}).

\bibitem[{\citenamefont{Brandt}(2005)}]{brandt05}
\bibinfo{author}{\bibfnamefont{E.~H.} \bibnamefont{Brandt}},
  \bibinfo{journal}{Phys. Rev. B} \textbf{\bibinfo{volume}{72}},
  \bibinfo{pages}{024529} (\bibinfo{year}{2005}).

\bibitem[{\citenamefont{L{\" o}rincz et~al.}(2004)\citenamefont{L{\" o}rincz,
  Welling, Rector, and Wijngaarden}}]{loerincz04}
\bibinfo{author}{\bibfnamefont{K.~A.} \bibnamefont{L{\" o}rincz}},
  \bibinfo{author}{\bibfnamefont{M.~S.} \bibnamefont{Welling}},
  \bibinfo{author}{\bibfnamefont{J.~H.} \bibnamefont{Rector}},
  \bibnamefont{and} \bibinfo{author}{\bibfnamefont{R.~J.}
  \bibnamefont{Wijngaarden}}, \bibinfo{journal}{Physica C}
  \textbf{\bibinfo{volume}{411}}, \bibinfo{pages}{1} (\bibinfo{year}{2004}).

\bibitem[{\citenamefont{Schuster et~al.}(1994)\citenamefont{Schuster, Indenbom,
  Koblischka, Kuhn, and Kronm{\" u}ller}}]{schuster94}
\bibinfo{author}{\bibfnamefont{T.}~\bibnamefont{Schuster}},
  \bibinfo{author}{\bibfnamefont{M.~V.} \bibnamefont{Indenbom}},
  \bibinfo{author}{\bibfnamefont{M.~R.} \bibnamefont{Koblischka}},
  \bibinfo{author}{\bibfnamefont{H.}~\bibnamefont{Kuhn}}, \bibnamefont{and}
  \bibinfo{author}{\bibfnamefont{H.}~\bibnamefont{Kronm{\" u}ller}},
  \bibinfo{journal}{Phys. Rev. B} \textbf{\bibinfo{volume}{49}},
  \bibinfo{pages}{3443} (\bibinfo{year}{1994}).

\bibitem[{\citenamefont{Gurevich}(2001)}]{gurevich01}
\bibinfo{author}{\bibfnamefont{A.}~\bibnamefont{Gurevich}},
  \bibinfo{journal}{Appl. Phys. Lett.} \textbf{\bibinfo{volume}{78}},
  \bibinfo{pages}{1891} (\bibinfo{year}{2001}).

\bibitem[{\citenamefont{Denisov et~al.}(2006)\citenamefont{Denisov, Shantsev,
  Galperin, Choi, Lee, Lee, Bobyl, Goa, Olsen, and Johansen}}]{denisov05}
\bibinfo{author}{\bibfnamefont{D.~V.} \bibnamefont{Denisov}},
  \bibinfo{author}{\bibfnamefont{D.~V.} \bibnamefont{Shantsev}},
  \bibinfo{author}{\bibfnamefont{Y.~M.} \bibnamefont{Galperin}},
  \bibinfo{author}{\bibfnamefont{E.-M.} \bibnamefont{Choi}},
  \bibinfo{author}{\bibfnamefont{H.-S.} \bibnamefont{Lee}},
  \bibinfo{author}{\bibfnamefont{S.-I.} \bibnamefont{Lee}},
  \bibinfo{author}{\bibfnamefont{A.~V.} \bibnamefont{Bobyl}},
  \bibinfo{author}{\bibfnamefont{P.~E.} \bibnamefont{Goa}},
  \bibinfo{author}{\bibfnamefont{A.~A.~F.} \bibnamefont{Olsen}},
  \bibnamefont{and} \bibinfo{author}{\bibfnamefont{T.~H.}
  \bibnamefont{Johansen}}, \bibinfo{journal}{Phys. Rev. Lett.}
  \textbf{\bibinfo{volume}{97}}, \bibinfo{pages}{077002}
  (\bibinfo{year}{2006}).

\bibitem[{\citenamefont{Prozorov et~al.}(2006)\citenamefont{Prozorov, Shantsev,
  and Mints}}]{prozorov}
\bibinfo{author}{\bibfnamefont{R.}~\bibnamefont{Prozorov}},
  \bibinfo{author}{\bibfnamefont{D.~V.} \bibnamefont{Shantsev}},
  \bibnamefont{and} \bibinfo{author}{\bibfnamefont{R.~G.} \bibnamefont{Mints}},
  \bibinfo{journal}{Phys. Rev. B} \textbf{\bibinfo{volume}{74}},
  \bibinfo{pages}{220511(R)} (\bibinfo{year}{2006}).

\bibitem[{\citenamefont{Shantsev et~al.}(2005)\citenamefont{Shantsev, Bobyl,
  Galperin, Johansen, and Lee}}]{shantsev05}
\bibinfo{author}{\bibfnamefont{D.~V.} \bibnamefont{Shantsev}},
  \bibinfo{author}{\bibfnamefont{A.~V.} \bibnamefont{Bobyl}},
  \bibinfo{author}{\bibfnamefont{Y.~M.} \bibnamefont{Galperin}},
  \bibinfo{author}{\bibfnamefont{T.~H.} \bibnamefont{Johansen}},
  \bibnamefont{and} \bibinfo{author}{\bibfnamefont{S.~I.} \bibnamefont{Lee}},
  \bibinfo{journal}{Phys. Rev. B} \textbf{\bibinfo{volume}{72}},
  \bibinfo{pages}{024541} (\bibinfo{year}{2005}).

\end{thebibliography}
\end{document}